\documentclass[twocolumn]{aastex63}
\usepackage{graphicx}
\usepackage{multirow}
\usepackage{longtable}
\usepackage{amsmath}

\shortauthors{Cai et al.}

\begin{document}

\title{Revisiting Approaches to Stellar White-Light Flare Energy Based on Spatiotemporally Resolved Solar Observations}

\author[0009-0007-4469-0663]{Yingjie Cai}
\affiliation{National Astronomical Observatories, Chinese Academy of Science, Beijing 100101, China}
\affiliation{School of Astronomy and Space Science, University of Chinese Academy of Sciences, Beijing 100049, China}
\affiliation{State Key Laboratory of Solar Activity and Space Weather, National Astronomical Observatories, Chinese Academy of Science, Beijing 100101, China}

\author[0000-0002-9534-1638]{Yijun Hou}
\affiliation{National Astronomical Observatories, Chinese Academy of Science, Beijing 100101, China}
\affiliation{School of Astronomy and Space Science, University of Chinese Academy of Sciences, Beijing 100049, China}
\affiliation{State Key Laboratory of Solar Activity and Space Weather, National Astronomical Observatories, Chinese Academy of Science, Beijing 100101, China}

\author[0000-0001-6655-1743]{Ting Li}
\affiliation{State Key Laboratory of Solar Activity and Space Weather, National Space Science Center, Chinese Academy of Sciences, Beijing 100190, China}
\affiliation{School of Astronomy and Space Science, University of Chinese Academy of Sciences, Beijing 100049, China}

\author[0000-0002-8258-4892]{Ying Li}
\affiliation{Key Laboratory of Dark Matter and Space Astronomy, Purple Mountain Observatory, Chinese Academy of Sciences, Nanjing 210023, China}
\affiliation{School of Astronomy and Space Science, University of Science and Technology of China, Hefei 230026, China}

\author[0000-0002-4978-4972]{Mingde Ding}
\affiliation{School of Astronomy and Space Science, Nanjing University, Nanjing 210023, China}
\affiliation{Key Laboratory of Modern Astronomy and Astrophysics (Nanjing University), Ministry of Education, Nanjing 210023, China}

\author[0000-0003-0057-6766]{Dechao Song}
\affiliation{Key Laboratory of Dark Matter and Space Astronomy, Purple Mountain Observatory, Chinese Academy of Sciences, Nanjing 210023, China}
\affiliation{School of Astronomy and Space Science, University of Science and Technology of China, Hefei 230026, China}

\author[0009-0006-3194-8525]{Changwen Zeng}
\affiliation{National Astronomical Observatories, Chinese Academy of Science, Beijing 100101, China}
\affiliation{School of Astronomy and Space Science, University of Chinese Academy of Sciences, Beijing 100049, China}

\author[0000-0002-2874-2706]{Jifeng Liu}
\affiliation{National Astronomical Observatories, Chinese Academy of Science, Beijing 100101, China}
\affiliation{School of Astronomy and Space Science, University of Chinese Academy of Sciences, Beijing 100049, China}
\affiliation{Institute for Frontiers in Astronomy and Astrophysics, Beijing Normal University, Beijing, China}
\affiliation{New Cornerstone Science Laboratory, National Astronomical Observatories, Chinese Academy of Sciences, Beijing, China}

\correspondingauthor{Yijun Hou}
\email{yijunhou@nao.cas.cn}

\begin{abstract}
Accurately estimating the bolometric energy of solar and stellar white-light flares (WLFs) is crucial for understanding their physical nature and impact on surrounding planets. However, the lack of spatial resolution in stellar observations forced pioneering stellar WLF studies to adopt simplified energy estimation methods, typically assuming either a constant flare temperature or a fixed radiating area. To assess the physical plausibility of these assumptions, we utilize high-spatiotemporal-resolution solar observations to analyze the true evolution of source region’s radiating area and temperature of 70 solar WLFs. It is revealed that both area and temperature of most solar WLFs undergo significant temporal evolution, and the flare area strongly correlates with the flare's peak optical continuum flux. Therefore, we propose a new energy estimation method that permits both flare area and temperature to evolve. Compared with existing methods, our dynamic approach yields systematically lower flare energies, which then prompts us to revisit classical macroscopic scaling laws related to the flare energy. It is further revealed that different energy estimation approaches can systematically alter these scaling relations, calling for a re‑examination of these established statistical results and their targeted testing or revision in future work. 
\end{abstract}


\section{Introduction}\label{sect1}
Solar flares are among the most energetic eruptive phenomena in the solar atmosphere, fundamentally driven by magnetic reconnection \citep{2002A&ARv..10..313P}. The rapidly released magnetic energy is converted into plasma thermal energy, bulk kinetic energy of the erupting filament and associated coronal mass ejection (CME), and kinetic energy of high-energy nonthermal particles \citep{2012ApJ...759...71E}. In the standard flare model, accelerated nonthermal electrons precipitate along field lines into the lower atmosphere and heat local dense plasma, which could drive the enhanced optical continuum emission observed as solar white-light flares \citep[WLFs]{2003A&A...403.1151D, 2008ApJ...688L.119J, 2018A&A...619A.100H,2026arXiv260510892H, 2020ApJ...904...96C, 2023ApJ...952L...6S, 2023ApJ...954....7L, 2024ApJ...963L...3L, 2025ApJ...979L..43Y, 2026RAA....26d7001C}. Analogous to these solar WLFs, numerous stellar flares have been observed on other stars by Kepler and the Transiting Exoplanet Survey Satellite (TESS) broadband optical data to release extraordinary energies of $10^{34}$ to $10^{38}$ erg, exceeding the most extreme recorded solar flares by several orders of magnitude \citep{2012Natur.485..478M, 2024Sci...386.1301V}. Such high energies of stellar WLFs are traditionally estimated from single-band photometric light curves using an isothermal blackbody approximation with a constant temperature (typically set to 10,000 K) \citep{2015SoPh..290.3663K, 2024LRSP...21....1K}. Under this assumption, the temporal evolution of the flare area is derived proportionally from the photometric light curve, and the bolometric energy is then calculated by integrating the Stefan‑Boltzmann emission flux ($\sigma_{\text{SB}} T_{\text{flare}}^4$) over area and time \citep{2013ApJS..209....5S}. 

Accurate determination of these bolometric energies has substantially advanced systematic statistical comparisons between solar and stellar flares, driving the quest for a unified physical paradigm \citep{2015EP&S...67...59M}. This pursuit rests on two key research directions. The first one focuses on the macroscopic energy distribution, i.e., the flare frequency distribution (FFD) \citep{1991SoPh..133..357H, 2012Natur.485..478M, 2013ApJS..209....5S, 2022Physi...5...11S, 2026ApJS..282...50Y}. \citet{2012Natur.485..478M} found that the occurrence frequency of stellar superflares follows a power-law distribution, suggesting a common physical mechanism for flares across a wide range of energies, and that the Sun could also produce superflares \citep{2015JPhCS.632a2058H, 2024Sci...386.1301V}. Subsequently, \citet{2013ApJS..209....5S} confirmed that the power-law index of the stellar FFD is approximately 2, consistent with solar flares \citep{1993SoPh..143..275C}. The second direction investigates the flare energy–duration relation. Magnetic reconnection theory predicts that the flare duration scales with the released energy as $\tau \propto E^{1/3}$, based on the Alfv{\'e}n transit time \citep{2015EP&S...67...59M, 2017ApJ...851...91N, 2024ApJ...975...69C}. \citet{2017ApJ...851...91N} revealed that the energy-duration relation on solar WLFs ($\tau \propto E^{0.38}$) is quite similar to that on stellar superflares ($\tau \propto E^{0.39}$). These consistent findings from the FFD and energy-duration relation indicate that stellar superflares, like their solar counterparts, arise from the release of stored magnetic energy through magnetic reconnection. Besides being essential for solar–stellar flare comparisons, accurate bolometric energy estimation of stellar WLFs is also crucial for assessing the impact of stellar eruptive activity on planetary habitability, e.g., severe atmospheric erosion and ozone depletion on orbiting exoplanets \citep{2018SciA....4.3302R, 2020IJAsB..19..136A}.

Despite its importance, accurate bolometric energy estimation for stellar WLFs from single-band photometric light curves remains highly contentious, primarily in two aspects. First, the radiative mechanism of the optical continuum emission of WLFs is still debated. While the isothermal blackbody approximation is widely used, recent studies suggest that the white-light (WL) emission may instead be dominated by an optically thin hydrogen free-bound (Hfb) continuum from the intensely heated chromosphere \citep{2014ApJ...783...98K, 2024MNRAS.532L..56H, 2024MNRAS.528.2562S, 2024LRSP...21....1K, 2026A&A...705A.157O}. In this model, flare energy deposited in the chromosphere ionizes predominant neutral hydrogen, producing a strong Hfb continuum. Second, regardless of the mechanism, flare energy estimation depends on two key independent physical variables: the flare radiating area and temperature, which cannot be solved simultaneously from a single equation (the stellar flare light curve) \citep{2003ApJ...597..535H, 2020ApJ...902..115H, 2025A&A...699A..90B}. To address this, researchers typically adopt either a variable-area model with an assumed constant temperature \citep{2013ApJS..209....5S, 2017ApJ...851...91N, 2024ApJ...975...69C}, or a variable-temperature model with a fixed area \citep{2025A&A...699A..90B, 2026ApJ...999L..18H}. The choice between these two energy estimation methodologies significantly affects the results. Using a variable-temperature model, \citet{2026ApJ...999L..18H} demonstrated that the traditional variable-area model with constant temperature can introduce systematic errors of up to an order of magnitude in bolometric energy estimates.

The lack of spatial resolution in stellar observations forced pioneering stellar WLF studies to adopt simplified energy estimation methods. While these approaches laid the crucial groundwork for understanding stellar WLFs, they struggle to accurately characterize the true evolution of the flare source region’s radiating area and temperature. Consequently, the artificial assumption of a constant temperature or a constant area in these approaches inevitably introduce substantial systematic biases into the final energy estimates \citep{2026RAA....26d7001C}. To resolve the parameter degeneracy and evaluate the biases in these existing energy estimation methods, spatially-resolved observations of WLFs are essential. In this study, using observations from the Helioseismic and Magnetic Imager (HMI) \citep{2012SoPh..275..207S} and the Atmospheric Imaging Assembly (AIA) \citep{2012SoPh..275...17L} aboard the Solar Dynamics Observatory (SDO) \citep{2012SoPh..275....3P}, we analyze the evolution of source region’s radiating area and temperature of 70 solar WLFs. Based on this, we propose a new energy estimation method with variable area and temperature that better matches the real evolution of solar WLFs, and systematically compare the results obtained by this new method with those from existing approaches.

\section{Observations}\label{sect2}
SDO/HMI provides solar full-disk photospheric continuum intensity maps via the Fe I 6173 {\AA} line, with a spatial pixel scale of $0.5^{\prime \prime}$ and a temporal cadence of 45 s. Through the improved WLF identification method introduced by \citet{2024ApJ...975...69C}, these HMI observations serve as the primary diagnostic for detecting localized WL emission enhancements during solar WLFs. However, here we must acknowledge an unavoidable concern: Fe I 6173 {\AA} line emission at flare peak could contaminate the HMI continuum channel, producing transient false signals that do not originate from true WL continuum brightening \citep{2017ApJ...839...67S}. To mitigate this, our identification method imposes dual requirements on signal duration and spatial clustering; through this procedure, the brief, point-like false signals caused by line contamination are systematically eliminated. Thus, despite the inherent limitations of the HMI pseudo-continuum, the WLF detections obtained with this optimized identification scheme remain relatively robust. And the HMI pseudo-continuum represents the best available compromise for studying the temporal evolution of WLF temperature and radiating area and for estimating bolometric energies under current observational conditions.

To ensure the robust derivation of physical parameters, we applied stringent selection criteria to initial candidates from two data sources, systematically excluding those with data gaps or complex background contamination. This screening ensures that only events with reliable optical continuum light curves are retained for our subsequent analyses. The finalized catalog comprises 70 solar WLFs: (1) 39 events (GOES class $\geq$ C5.0) recorded from February 2011 to July 2023 \citep{2024ApJ...975...69C}; and (2) 31 recent solar WLFs (GOES class $\geq$ M1.0) produced by the super active region NOAA 13664/13697 from 2 May to 9 June 2024 (Cai et al. 2026, in preparation). Ultimately, this sample consists of 33 X-class, 32 M-class, and 5 C-class solar WLFs.

\section{Existing WLF Energy Estimation Methods}\label{sect3}
For the calculation of bolometric energy of stellar WLFs, the typical observations available are single-band light curves. In practice, these raw single-band light curves are usually detrended and normalized to extract the flare amplitude $C(t)=\frac{L - L_{\text{star}}}{L_{\text{star}}}$, which directly reflects the relationship between the observed total luminosity during WLFs ($L$) and the quiescent stellar luminosity ($L_{\text{star}}$). The observed total luminosity $L$ is described by:
\begin{equation}
L = (A_{\text{star}} - A_{\text{flare}})F_{\text{star}} + A_{\text{flare}}F_{\text{flare}}, \label{eq:1}
\end{equation}
where $F_{\text{flare}}$ and $F_{\text{star}}$ represent the observed radiative fluxes of the flaring region and the quiescent star, $A_{\text{flare}}$ and $A_{\text{star}}$ denote the areas of the flare and the stellar disk. The observed luminosity of the quiescent star is:
\begin{equation}
L_{\text{star}} = A_{\text{star}}F_{\text{star}}. \label{eq:2}
\end{equation}
The flare amplitude $C(t)$ is then given by:
\begin{equation}
C(t) = \frac{A_{\text{flare}} \times (F_{\text{flare}} - F_{\text{star}})}{A_{\text{star}} \times F_{\text{star}}}. \label{eq:3}
\end{equation}
The observed radiative flux $F$ is obtained by integrating the specific intensity $I_{\lambda}$ over both the wavelength range and the solid angle of the emitting hemisphere: 
\begin{multline}
F = \int_{\lambda_1}^{\lambda_2} S_{\lambda} I_{\lambda} \left[ \int_0^{2\pi} d\phi \int_0^{\pi/2} \cos\theta \sin\theta d\theta \right] d\lambda \\
= \pi \int_{\lambda_1}^{\lambda_2} S_{\lambda} I_{\lambda} d\lambda, \label{eq:4}
\end{multline}
where $S_{\lambda}$ is the response function of the instrument. In characterizing the macroscopic emission of the quiescent star ($F_{\text{star}}$), the specific intensity $I_{\lambda}$ is typically modeled by a Planck function $B_{\lambda}(T_{\text{star}})$ corresponding to the stellar effective temperature $T_{\text{star}}$. 

The total luminosity of any emitting source is simply the product of its radiating area and its surface flux. Thus, the flare luminosity $L_{\text{flare}}(t)$ can be found by:
\begin{equation}
L_{\text{flare}}(t) = A_{\text{flare}}(t) \pi \int_0^{\infty} I_{\lambda}(T_{\text{flare}}(t)) d\lambda, \label{eq:5}
\end{equation}
where $T_{\text{flare}}(t)$ represent the temperature of the flare. The total bolometric energy of flare ($E_{\text{bol}}$) is then given by the temporal integration of the flare luminosity:
\begin{multline}
E_{\text{bol}} = \int_{t_{\text{start}}}^{t_{\text{end}}} L_{\text{flare}}(t) dt \\ 
= \int_{t_{\text{start}}}^{t_{\text{end}}} A_{\text{flare}}(t) \left(\pi \int_0^{\infty} I_{\lambda}(T_{\text{flare}}(t)) d\lambda \right) dt. \label{eq:6}
\end{multline}

It is obvious that $E_{\text{bol}}$ depends on two key variables: $A_{\text{flare}}(t)$ and $T_{\text{flare}}(t)$, which however cannot be solved simultaneously from a single observable $C(t)$. To circumvent this parameter degeneracy, various methodologies adopting distinct physical assumptions have been developed as follows.

\subsection{Blackbody Model with Fixed Temperature and Variable Area}\label{sect3.1}
This traditional approach, widely adopted in statistical studies of stellar superflares, models the flare continuum as optically thick blackbody radiation with a temporally invariant effective temperature, typically fixed at $T_{\text{flare}} = 10,000$ K \citep{2013ApJS..209....5S, 2015SoPh..290.3663K, 2024LRSP...21....1K}. 
Moreover, assuming the flaring region is significantly smaller than the stellar disk ($A_{\text{flare}} \ll A_{\text{star}}$), from Equations \ref{eq:3} and \ref{eq:4}, the temporal evolution of the flare area is deduced directly from the flare amplitude $C(t)$:
\begin{equation}
A_{\text{flare}}(t) = C(t) \pi R_{\text{star}}^2 \frac{\int S_{\lambda} B_{\lambda}(T_{\text{star}}) d\lambda}{\int S_{\lambda} B_{\lambda}(T_{\text{flare}}) d\lambda}, \label{eq:7}
\end{equation}
where $R_{\text{star}}$ is the stellar radius, and the specific intensity $I_{\lambda}(T_{\text{flare}})$ is replaced by the Planck function $B_{\lambda}(T_{\text{flare}})$. Because $T_{\text{flare}}$ is strictly held constant, the derived flare area serves as the sole temporal variable. 

Under the assumption of blackbody radiation, the total flare flux satisfies the Stefan-Boltzmann law: 
\begin{equation}
F = \pi \int_0^{\infty} B_{\lambda}(T_{\text{flare}}) d\lambda = \sigma_{\text{SB}} T_{\text{flare}}^4, \label{eq:8}
\end{equation}
where $\sigma_{\mathrm{SB}}$ represents the Stefan-Boltzmann constant. Finally, based on Equation \ref{eq:6}, the bolometric energy can be obtained by integrating the Stefan-Boltzmann emission flux over the entire flare duration: 
\begin{equation}
E_{\text{bol, 1}} = \sigma_{\text{SB}} T_{\text{flare}}^4 \int_{t_{\text{start}}}^{t_{\text{end}}} A_{\text{flare}}(t) dt. \label{eq:9}
\end{equation}

\subsection{Hydrogen Free–bound Continuum Model with Fixed Temperature and Variable Area}\label{sect3.2}
Some studies argued that the optical continuum of solar flares may not originate from an optically thick photosphere, but rather from an optically thin Hfb continuum formed in the intensely heated chromosphere \citep{2014ApJ...783...98K, 2024MNRAS.532L..56H}. As a result, the specific intensity $I_{\lambda}(T_{\text{flare}})$ in Equations \ref{eq:4} and \ref{eq:5} is replaced by the Hfb specific intensity $I_{\lambda}^{\text{Hfb}}$ \citep{2024MNRAS.528.2562S}. Assuming an isothermal, uniform slab of chromospheric plasma \citep{2014ApJ...783...98K}, this specific intensity is analytically given by:
\begin{multline}
I_{\lambda}^{\text{Hfb}} = \frac{6.48 \times 10^{-14}}{4\pi \lambda^2} \frac{T_{\text{flare}}^{-3/2}}{n^3} \\
\times \exp\left( \frac{1.58 \times 10^5}{n^2 T_{\text{flare}}} - \frac{1.44 \times 10^8}{\lambda T_{\text{flare}}} \right) n_e^2 L, \label{eq:10}
\end{multline}
where $n$ is the principal quantum number for the recombining level, and $n_e^2 L$ is the column emission measure. Following standard empirical calibrations, we adopt $T_{\text{flare}} = 10,000$ K and $n_e^2 L = 7 \times 10^{34}~\text{cm}^{-5}$ \citep{2011A&A...530A..84K, 2014ApJ...783...98K}. Similar to Equation \ref{eq:7}, the temporal evolution of the flare radiating area becomes:
\begin{equation}
A_{\text{flare}}(t) = C(t) \pi R_{\text{star}}^2 \frac{\int S_{\lambda} B_{\lambda}(T_{\text{star}}) d\lambda}{\int S_{\lambda} I_{\lambda}^{\text{Hfb}}(T_{\text{flare}}) d\lambda}. \label{eq:11}
\end{equation}
Based on Equation \ref{eq:6}, the total bolometric energy under the Hfb continuum model ($E_{\text{bol, 2}}$) is then calculated by integrating the wavelength-integrated Hfb continuum over the entire flare duration:
\begin{equation}
E_{\text{bol, 2}} = \pi \int_{t_{\text{start}}}^{t_{\text{end}}} A_{\text{flare}}(t)  \int_0^{\infty} I_{\lambda}^{\text{Hfb}}(T_{\text{flare}}) d\lambda dt. \label{eq:12}
\end{equation}

\subsection{Blackbody Model with Fixed Area and Variable Temperature}\label{sect3.3}
In contrast to the above approaches with a fixed temperature and a variable area, some recent studies suggested that the spatial extent of the flaring region remains relatively stable, while the temperature evolves significantly \citep{2023A&A...673A.150C, 2024A&A...690A.254G}. To better capture this behavior, \citet{2026ApJ...999L..18H} built an alternative blackbody framework assuming a constant flare area and numerically resolved the time-dependent temperature of stellar WLFs. Without assuming $A_{\text{flare}} \ll A_{\text{star}}$, based on Equations \ref{eq:3} and \ref{eq:4}, the authors derived the flare area from the flare amplitude $C(t)$: 
\begin{equation}
A_{\text{flare}} = \frac{C(t) \pi R_{\text{star}}^2 \int S_{\lambda} B_{\lambda}(T_{\text{star}}) d\lambda}{\int S_{\lambda} B_{\lambda}(T_{\text{flare}}(t)) d\lambda - \int S_{\lambda} B_{\lambda}(T_{\text{star}}) d\lambda}. \label{eq:13}
\end{equation}
In this method, the constant flare area $A_{\text{flare}}$ is anchored at flare peak via the empirically constrained maximum temperature $T_{\text{peak}}$ and the maximum flare amplitude $C_{\text{peak}}$. Since extrapolating $T_{\mathrm{peak}}$ from semi-empirical grids is currently unfeasible for the Sun \citep{2025A&A...699A..90B}, here we follow \citet{2026ApJ...999L..18H} and adopt a traditional $T_{\mathrm{peak}} = 10,000$ K, thereby enabling a direct, controlled comparison with the above two fixed-temperature approaches. Once $A_{\text{flare}}$ is fixed, based on Equation \ref{eq:13}, the dynamic temperature $T_{\text{flare}}(t)$ is determined at each observational time step by solving the integral equation:
\begin{multline}
\int S_{\lambda} B_{\lambda}(T_{\text{flare}}(t)) d\lambda \\
= \left[ C(t) \frac{\pi R_{\text{star}}^2}{A_{\text{flare}}} + 1 \right] \int S_{\lambda} B_{\lambda}(T_{\text{star}}) d\lambda. \label{eq:14}
\end{multline}
Finally, based on Equations \ref{eq:6} and \ref{eq:8}, the bolometric energy is calculated by integrating the varying luminosity over the flare duration:
\begin{equation}
E_{\text{bol, 3}} = \sigma_{\text{SB}} A_{\text{flare}} \int_{t_{\text{start}}}^{t_{\text{end}}} T_{\text{flare}}^4(t) dt. \label{eq:15}
\end{equation}

\section{Results and Discussion}\label{sect4}
\subsection{True Evolution Characteristics of Radiating Area and Temperature in Solar WLFs}\label{sect4.1}

\begin{figure*}[htbp]
\centering
\includegraphics [width=0.99\textwidth]{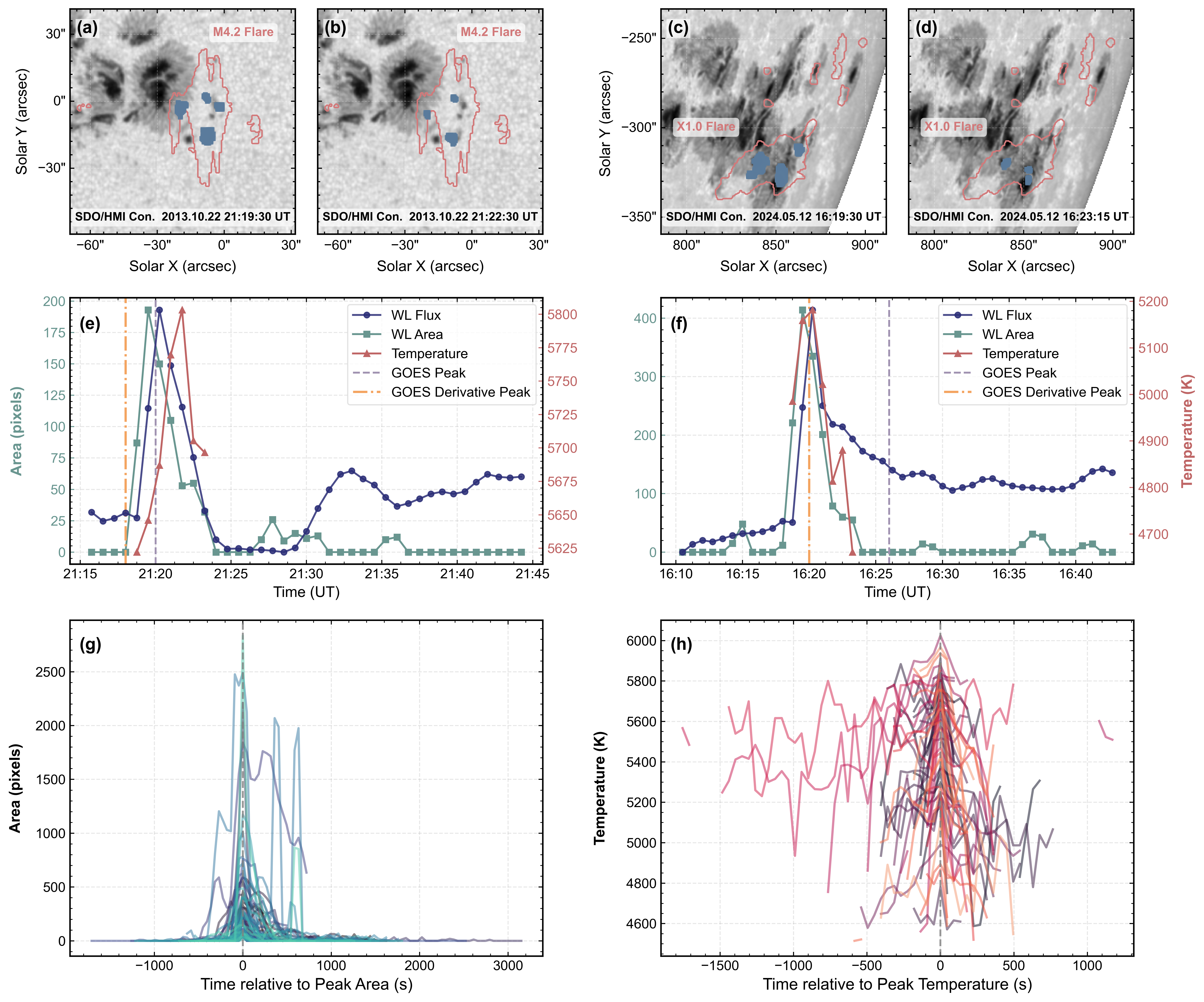}
\caption{Evolution of the integrated WL flux, flare radiating area, and temperature for solar WLFs. (a) and (b) Spatial distributions of WL emission enhancement signals (blue) at different times during an M4.2 solar WLF. The red contour encompasses the flare ribbon region in AIA 1600 {\AA} observations. (c) and (d) Similar to (a) and (b), but for an X1.0 solar WLF. (e) and (f) The temporal evolution of the integrated WL flux, flare area, and temperature for the two representative WLFs. Purple and orange vertical dashed lines indicate the peak times of GOES SXR 1--8 {\AA} flux and its derivative, respectively. (g) and (h) The temporal evolution of the flare area and temperature for the other 68 WLFs.}
\label{Fig1}
\end{figure*}

To assess the physical plausibility of the existing approaches to stellar WLF energy assuming a fixed flare temperature or radiating area, we first investigate the true evolution of source region’s radiating area and temperature of 70 solar WLFs based on spatiotemporally resolved solar observations. Following \citet{2024ApJ...975...69C}, we identify WL emission enhancements at each time step of these WLFs; the instantaneous radiating area is obtained from the spatial integral of these signals, and the flare temperature is then calculated as the unweighted spatial average of the flaring pixels' effective blackbody temperatures derived from their WL intensities under a blackbody approximation (see more details in Sec. 3.4.2 of \citet{2026RAA....26d7001C}). Figure \ref{Fig1} presents evolution of the calculated area and temperature of two representative events and the other 68 solar WLFs. For the M4.2 flare shown in panels (a), (b), and (e), its flare area exhibits obvious temporal variation, which strongly correlates with the optical continuum light curve. Its temperature also varies during the flare. The X1.0 solar WLF shown in panels (c), (d), and (f) exhibits similar evolution of area and temperature. Figures \ref{Fig1}(g) and (h) further demonstrate that the majority of the remaining 68 WLFs exhibit pronounced temporal variations in both area and temperature.

\begin{figure*}[htbp]
\centering
\includegraphics [width=0.95\textwidth]{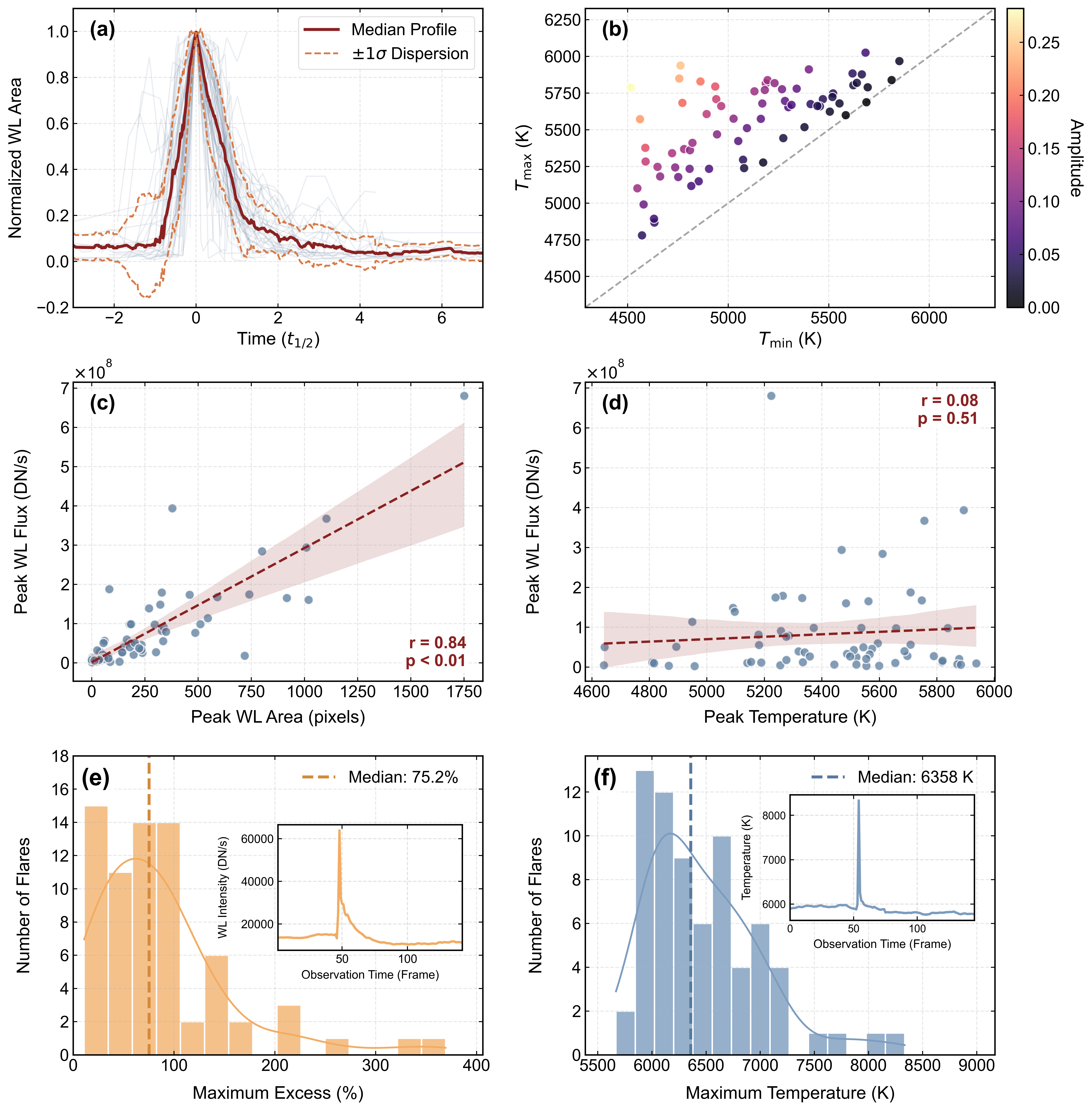}
\caption{Statistical analysis of the integrated WL flux, flare radiating area, and temperature across the 70 solar WLFs. (a) Flare area evolution template of 70 solar WLFs, scaled to relative time ($t_{1/2}$, the full width at half maximum) and amplitude. The median of all 70 solar WLFs (red solid line), as well as the robust standard deviation (orange dotted lines), are overlaid. (b) Scatter plot of maximum versus minimum temperatures, with colors indicating the maximum amplitude of temperature variations during flares. (c) and (d) Correlations between peak WL flux and peak flare area or peak temperature. The dark red dashed lines represent the linear regression fits to the data, while the corresponding shaded bands indicate the 95\% confidence intervals of the fits. The Pearson correlation coefficient ($r$) and the statistical significance ($p$-value) for each relationship are annotated within the respective panels. (e) Distribution of the maximum intensity excess among all pixels for each WLF, with a median value of 75.2\%. The inset displays the intensity evolution of the pixel with the largest maximum intensity excess across all flares. (f) Similar to (e), but for the maximum temperature with a median value of 6358 K.}
\label{Fig2}
\end{figure*}

Furthermore, as shown in Figure \ref{Fig2}(a), the flare area evolution template extracted from the sample of 70 solar WLFs further corroborates the highly dynamic evolution of the radiating area during flares. Figure \ref{Fig2}(b) shows that the flare temperature also evolves with the maximum amplitude of 28.2\%, but its relative variation is moderate compared to the area. Crucially, we correlate peak WL flux of WLF with peak area and peak temperature (Figures \ref{Fig2}(c) and (d)) and find a robust positive correlation (0.84) with area, but no significant correlation (0.08) with temperature. Finally, a pixel-level analysis of all the 70 WLFs reveals that the maximum intensity excess and peak temperature have median values of 75.2\% and 6358 K, respectively, and the largest values of 369.3\% and 8334 K, respectively (Figures \ref{Fig2}(e) and (f)). 

These findings, based on solar WLF observations, effectively challenge the existing WLF energy estimation models assuming either a constant flare temperature or a fixed radiating area. The observed pronounced temporal evolution of the flare temperature directly contradicts the assumption of a constant flare temperature. The strong correlation between WL flux and flare area indicates that the expansion of the flare area plays a crucial role in the overall radiative enhancement of WLFs, contradicting the fixed-area assumption. This calls for a new WLF energy estimation approach that adapts to these newly revealed observational constraints that both flare area and temperature can evolve.

\subsection{A New Energy Estimation Method with Variable Area and Temperature}\label{sect4.2}
Here, we propose a new energy estimation method with variable flare area and time-dependent and pixel-dependent temperature, that better matches the true evolution characteristics of solar WLFs. Based on the following two facts: (1) the WLF emission is physically a combination of optically thin Hfb continuum emission ($I_{\lambda}^{\text{Hfb}}$) originating from the chromospheric condensation and optically thick blackbody emission ($B_{\lambda}$) from the underlying flare-heated photosphere \citep{2014ApJ...783...98K}; (2) for narrowband observations at a specific wavelength, the instrument response function ($S_{\lambda}$) in Equation \ref{eq:4} can be factored out \citep{2026RAA....26d7001C}, we obtain the following equation:
\begin{multline} 
\frac{I_{\text{flare}}(x,y,t)}{I_{\text{bg}}(x,y)} = \\
\frac{I_{\lambda=6173 \mathring{A}}^{\text{Hfb}}(T_{\text{chromo}}(x,y,t)) + B_{\lambda=6173 \mathring{A}}(T_{\text{photo}}(x,y,t))}{B_{\lambda=6173 \mathring{A}}(T_{\text{bg}}(x,y))}, \label{eq:16}
\end{multline}
where $I_{\text{flare}}(x,y,t)$ and $I_{\text{bg}}(x,y)$ are the WL intensity of a flaring pixel and its pre-flare background intensity observed by HMI. $T_{\text{chromo}}(x,y,t)$ represents the optically thin chromospheric condensation region temperature of a flaring pixel and $T_{\text{photo}}(x,y,t)$ denotes its effective temperature of the flare-heated underlying photosphere. The corresponding quiescent temperature before flare ($T_{\text{bg}}(x,y)$) is empirically determined for each flaring pixel based on its specific location, which can be categorized into three regions with assigned temperatures: umbra ($4500$ K), penumbra ($5500$ K), and quiet Sun ($5780$ K). 

Subsequently, we need to derive the flare temperature ($T_{\text{chromo}}(x,y,t)$ and $T_{\text{photo}}(x,y,t))$) for every pixel at each time step for eventual energy estimation. However, given that HMI pseudo-continuum observations provide only one observational constraint, we cannot simultaneously resolve these two temperature variables. To bypass this problem, we can adopt two simplified scenarios. The first one assumes a strictly photospheric origin, neglecting the chromospheric contribution. This is physically reasonable under mechanisms such as direct precipitation of highly energetic proton beams into the lower atmosphere \citep{2018ApJ...862...76P, 2019SoPh..294..103T, 2020SoPh..295..174M}, magnetic reconnection in the lower atmosphere \citep{2020ApJ...893L..13S}, or Alfv{\'e}n wave dissipation in the photosphere \citep{2008ApJ...675.1645F, 2025ApJ...986L..15X, 2025ApJ...987..163T}. Under this assumption, the flare emission is treated as an optically thick blackbody, which aligns with the traditional method \citep{2013ApJS..209....5S}. Equation \ref{eq:16} can be rearranged to yield:
\begin{equation} 
\frac{I_{\text{flare}}(x,y,t)}{I_{\text{bg}}(x,y)} = \frac{B_{\lambda=6173 \mathring{A}}(T_{\text{flare}}(x,y,t))}{B_{\lambda=6173 \mathring{A}}(T_{\text{bg}}(x,y))}. \label{eq:17}
\end{equation}
By inverting this relation, $T_{\text{flare}}(x,y,t)$ can be explicitly obtained for each spatially resolved flaring pixel at every time step. 

\begin{figure*}[htbp]
\centering
\includegraphics [width=0.85\textwidth]{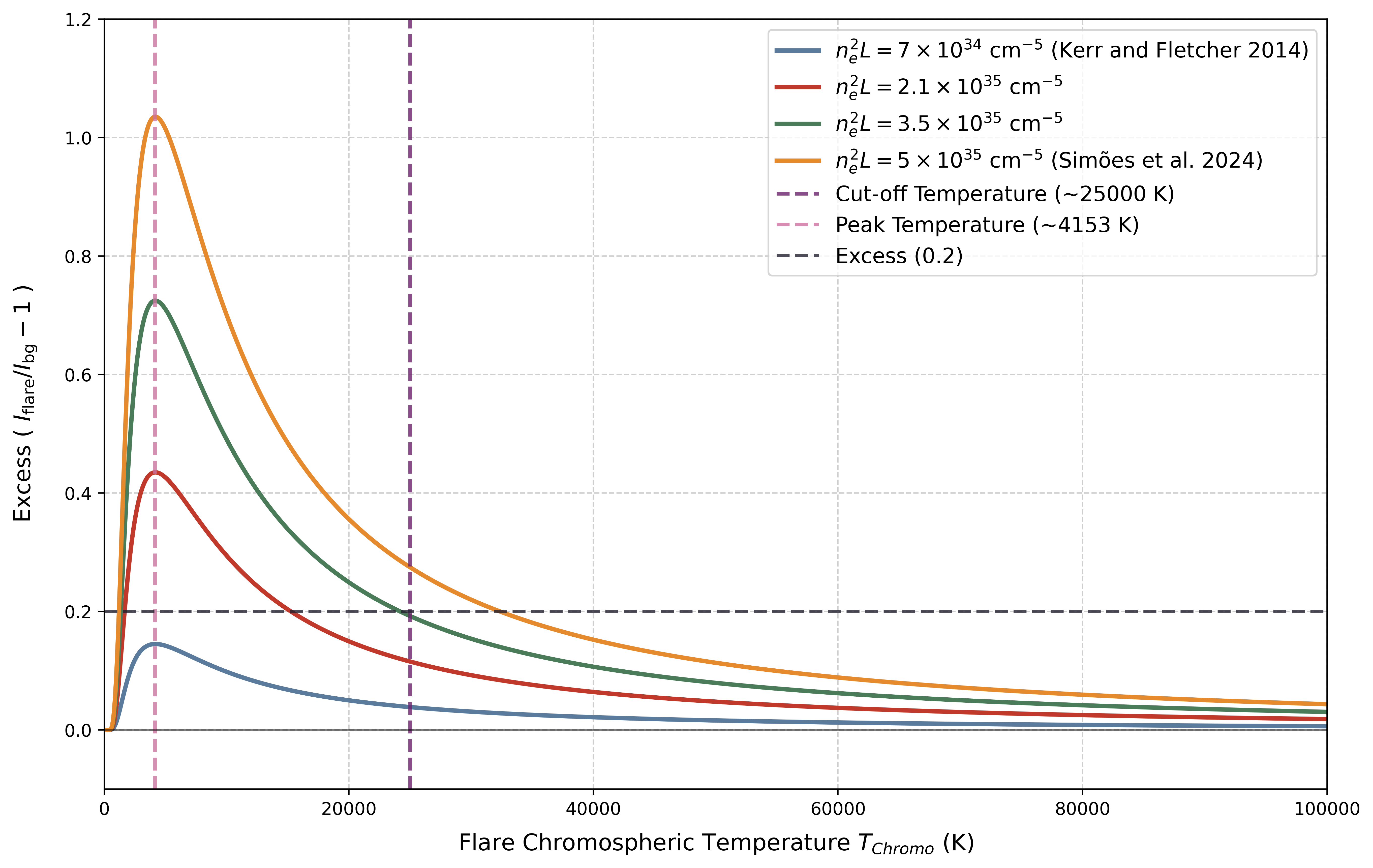}
\caption{Theoretical evolution of the flare-to-background intensity excess as a function of chromospheric temperature under varying column emission measures ($n_{\text{e}}^2L$). The four solid curves illustrate the results for different $n_{\text{e}}^2L$, ranging from $7 \times 10^{34} \ \mathrm{cm}^{-5}$ to $5 \times 10^{35} \ \mathrm{cm}^{-5}$. The vertical pink dashed line indicates the peak ratio, which consistently occurs at $\sim 4153 \ \mathrm{K}$. The vertical purple dashed line marks the physical cut-off temperature at $\sim 25,000 \ \mathrm{K}$. The horizontal grey dashed line represents a flare-to-background intensity excess of 0.2.}
\label{Fig3}
\end{figure*}

The second scenario assumes a strictly chromospheric origin, where the optically thin $I_{\text{Hfb}}$ emission superimposes on the unperturbed photospheric background \citep{2014ApJ...783...98K}. Equation \ref{eq:16} can be rewritten as: 
\begin{multline} 
\frac{I_{\text{flare}}(x,y,t)}{I_{\text{bg}}(x,y)}-1 = \frac{I_{\lambda=6173 \mathring{A}}^{\text{Hfb}}(T_{\text{chromo}}(x,y,t))} {B_{\lambda=6173 \mathring{A}}(T_{\text{bg}}(x,y))}. \label{eq:18}
\end{multline}
Assuming $T_{\text{bg}}$ as 5780 K, functional relationship between the observed flare-to-background intensity \textbf{excess ($\frac{I_{\text{flare}}}{I_{\text{bg}}}-1$)} and flaring chromospheric temperature ($T_{\text{chromo}}$) is shown in Figure \ref{Fig3}. One can see that deriving $T_{\text{chromo}}$ from this relation will encounter three insurmountable problems. (1) First, to obtain a temperature solution, the observed intensity \textbf{excess must be greater than 0} and less than the maximum of the function curve, which however is frequently unmet in the actual solar WLF observations. For example, observed intensity excess of 75.2\% (see Figure \ref{Fig2}(e)) can easily surpass the peak of the blue, red, and green curves in Figure \ref{Fig3}, yielding no solution. (2) Second, even when the first condition above is met, the inferred temperature is not unique. For a fixed column emission measure ($n_{\text{e}}^2L$), the same excess can correspond to two temperature solutions. The low-temperature solution ($T < 4153\text{ K}$) is discarded because it contradicts actual flaring chromosphere conditions. The high-temperature solution could be higher than the cut-off temperature (25,000 K) of the Hfb recombination model for a modest intensity enhancement. For instance, fitting a 20\% excess under the orange curve forces $T > 30,000\text{ K}$, where hydrogen is nearly fully ionized and the Hfb recombination model breaks down. (3) More importantly, as shown by Equation \ref{eq:10} and Figure \ref{Fig3}, $I_{\lambda}^{\text{Hfb}}$ is much more sensitive to $n_{\text{e}}^2L$ than to $T_{\text{chromo}}$. Therefore, if the Hfb recombination model is adopted, $n_{\text{e}}^2L$ should be regarded as the primary diagnostic quantity rather than $T_{\text{chromo}}$. As illustrated in Figure \ref{Fig3}, changing $n_{\text{e}}^2L$ produces a substantial vertical shift of the excess profile. The adopted values of $n_{\text{e}}^2L$ are $7\times10^{34}$, $2.1\times10^{35}$, $3.5\times10^{35}$, and $5\times10^{35}~{\rm cm^{-5}}$ for the blue, red, green, and orange curves, respectively. In reality, $n_{\text{e}}^2L$ is highly dynamic \citep{2017A&A...605A.125S, 2023A&A...673A.150C}: using a uniform value is inappropriate even for different pixels of the same WLF kernel, let alone for different WLFs. Therefore, without an independent constraint on $n_{\text{e}}^2L$, the Hfb recombination model cannot provide a robust temperature diagnostic from the observed intensity excess alone.

\begin{figure*}[htbp]
\centering
\includegraphics [width=0.9\textwidth]{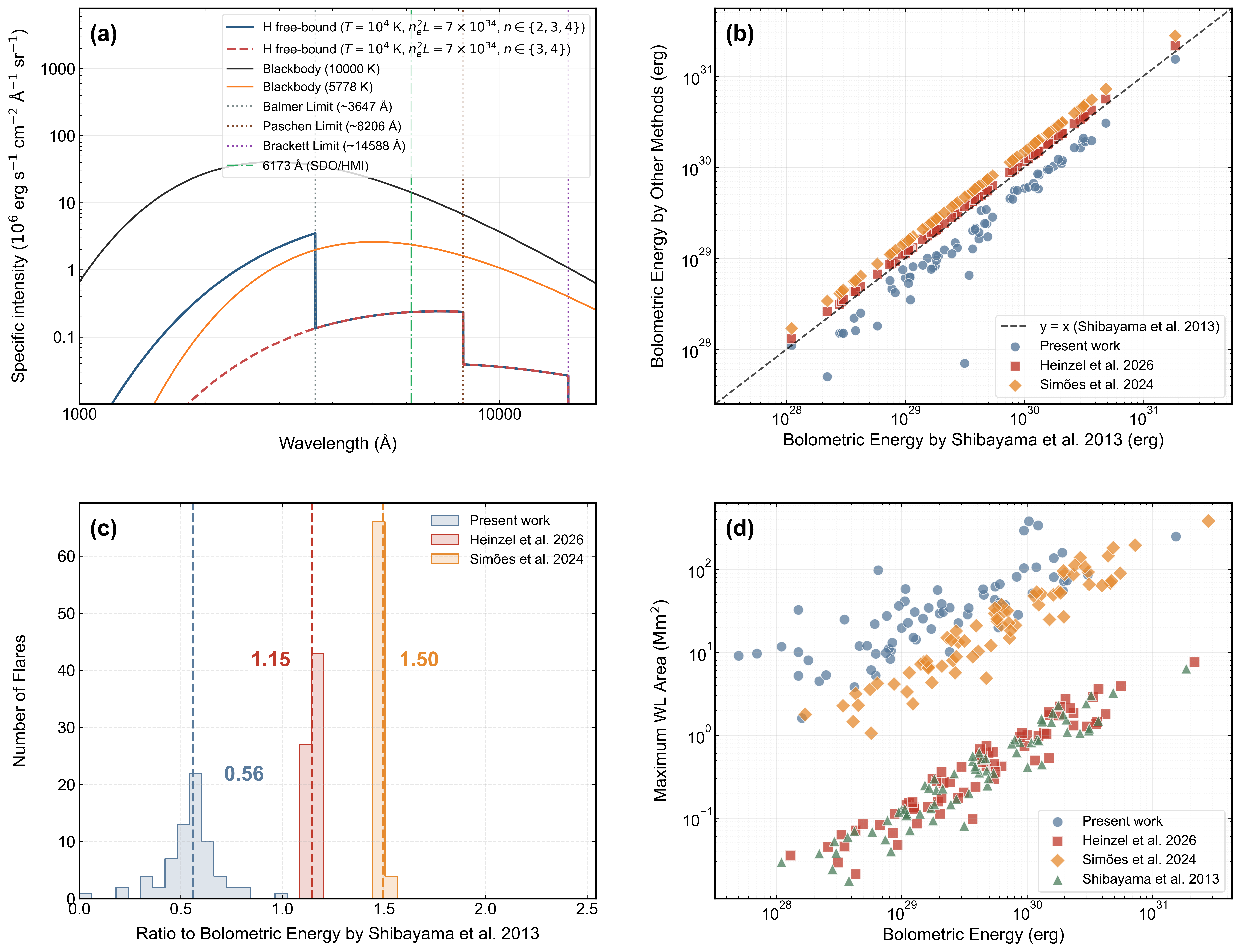}
\caption{Comparison of energy estimates from different methods. (a) Specific intensity spectra for the blackbody model and Hfb continuum model. (b) and (c) Scatter plot and histogram comparing the bolometric energy derived from the present work and another two methods \citep{2024MNRAS.528.2562S, 2026ApJ...999L..18H} with respect to the traditional blackbody model with fixed temperature \citep{2013ApJS..209....5S}. (d) Maximum flare area versus bolometric energy estimated from these methods.}
\label{Fig4}
\end{figure*}

As a result, we ultimately adopted the first scenario as our temperature computational framework for the new energy estimation method. Nevertheless, we must emphasize that although the hydrogen recombination continuum model is unsuitable for pixel-level temperature inversion, it remains effective for deriving the flare temperature through the spatially-integrated light curve, because the intensity enhancement of the entire flaring region typically falls within a mathematically solvable range \citep{2024MNRAS.528.2562S}. Utilizing the temperatures retrieved under the first scenario, the bolometric luminosity is accurately reconstructed according to Equations \ref{eq:5} and \ref{eq:8} by integrating the Stefan-Boltzmann emission exclusively over the dynamically evolving flare area $S(t)$:
\begin{equation}
L_{\text{bol}}(t) = \int_{S(t)} \sigma_{\text{SB}} T_{\text{flare}}^4(x,y,t) dS. \label{eq:19}
\end{equation}
After subtracting the quiescent background, the absolute bolometric energy of the flare is computed by integrating the excess bolometric luminosity over the flare duration:
\begin{equation}
E_{\text{bol, 4}} = \int_{t_{\text{start}}}^{t_{\text{end}}} \left[ L_{\text{bol}}(t) - L_{\text{bg}}(t) \right] dt. \label{eq:20}
\end{equation}
This new energy estimation method permits both flare area and temperature to evolve, well matching the true evolution characteristics of solar WLFs.

\subsection{Systematic Comparison of Bolometric Energy Estimates from Different Methods}\label{sect4.3}

Figure \ref{Fig4} presents a comprehensive comparison of bolometric energy estimates of 70 solar WLFs from different methods introduced by Sections \ref{sect3} and \ref{sect4.2}. Figures \ref{Fig4}(b) and (c) show that, compared to the classic blackbody model with fixed temperature \citep{2013ApJS..209....5S}, our method calculates smaller flare energy, whereas the other two methods \citep{2024MNRAS.528.2562S,2026ApJ...999L..18H} yield relatively larger energies. The median energy ratios of our method, \citet{2026ApJ...999L..18H}, and \citet{2024MNRAS.528.2562S} with respect to \citet{2013ApJS..209....5S} are 0.56, 1.15, and 1.50, respectively. Furthermore, our method recovers the largest maximum flare areas, followed by the Hfb continuum model \citep{2024MNRAS.528.2562S}, whereas the two optically thick blackbody models \citep{2013ApJS..209....5S,2026ApJ...999L..18H} yield significantly smaller and comparable areas (Figure \ref{Fig4}d).

\begin{figure*}[htbp]
\centering
\includegraphics [width=0.99\textwidth]{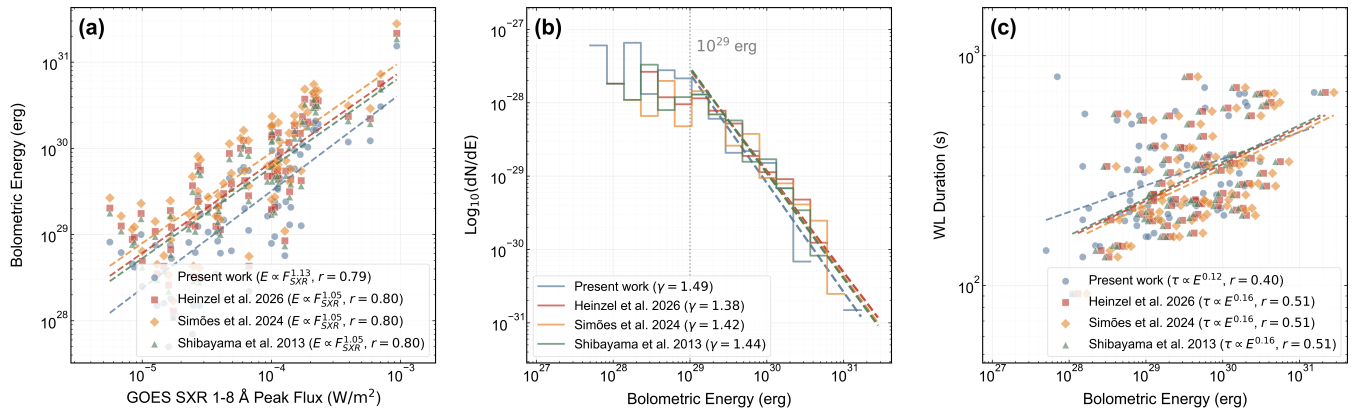}
\caption{Impact of different energy estimation methodologies on macroscopic flare scaling laws. (a) Scaling relationship between bolometric energy and GOES SXR 1--8 {\AA} peak flux. (b) Flare frequency distribution (FFD) of the bolometric energy. (c) Relationship between bolometric energy and flare duration. The dashed lines represent the best power-law fits corresponding to the different datasets and energy estimation methodologies. $r$ and $\gamma$ denote the correlation coefficient of the fitting and derived power-law index, respectively.}
\label{Fig5}
\end{figure*}

The systematic energy discrepancies among these methodologies originate from their underlying physical assumptions. As illustrated in Figure \ref{Fig4}(a), the specific intensity of the optically thin Hfb continuum at the SDO/HMI observing wavelength ($6173~\text{\AA}$, green dash-dotted line) is substantially weaker than that of a $10,000$ K optically thick blackbody. For the Hfb continuum model \citep{2024MNRAS.528.2562S} versus the traditional blackbody model \citep{2013ApJS..209....5S}, the energy ratio can be analytically derived to be a constant ($\sim1.50$). For the blackbody model with variable temperature \citep{2026ApJ...999L..18H}, the scaling factor relative to the traditional blackbody model depends implicitly on the light-curve morphology of flares; however, the similarity of solar flare light-curve morphology \citep{2021MNRAS.502.3922K} confines this factor to a narrow range, yielding clustered ratios ($\sim1.15$). Our method gives the lowest energies, which is a direct consequence of the extreme temperature sensitivity of the Stefan–Boltzmann law: since the radiative output scales as $T^4$, even a moderate temperature drop sharply reduces the inferred energy. The temperatures derived by our method remain well below $10,000$ K, which naturally results in systematically smaller energy estimates. To test this, we lowered the fixed temperature used in the traditional method \citep{2013ApJS..209....5S} from 10,000 K to 5,500 K, the average peak temperature of solar WLFs obtained by our method. With this modification, the flare energy ratio (our method vs. traditional method) changed from 0.56 to 0.83, proving that the temperature setting plays a crucial role in determining the estimated energy.

\subsection{Revisiting Energy Scaling Laws of Solar/Stellar WLFs}\label{sect4.4}

To evaluate how the systematic energy deviations between different methods uncovered in Section \ref{sect4.2} propagate into macroscopic stellar physics, we contrast several fundamental scaling laws involved solar/stellar flare energy based on the energy estimates of 70 solar WLFs from these methods. As shown in Figure \ref{Fig5}(a), all methods show a robust positive correlation between bolometric energy and peak soft X-ray (SXR) flux, but the existing models yield $E \propto F_{\mathrm{SXR}}^{1.05}$, whereas ours gives a steeper $E \propto F_{\mathrm{SXR}}^{1.13}$ with a systematic downward energy shift. For the FFD of flare energy (Figure \ref{Fig5}(b)), the existing models yield shallow power-law indices ($\gamma \approx 1.38\text{--}1.44$), whereas the energies derived by our method generate a little steeper high-energy tail ($\gamma = 1.49$). Finally, although all frameworks confirm that more energetic flares last longer (Figure \ref{Fig5}(c)), our method yield a notably shallower energy-duration scaling ($\tau \propto E^{0.12}$) than that of other methods ($\tau \propto E^{0.16}$).

The above small deviations in scaling-law index could introduce large systematic biases in statistical extrapolations. Empirical SXR–energy relations are widely used to estimate bolometric energy of solar flares \citep{2010NatPh...6..690K, 2012ApJ...759...71E}, yet our results reveal that methodological choices critically distort these relations. Furthermore, FFD-based occurrence rate predictions of solar superflares \citep{2012Natur.485..478M, 2024Sci...386.1301V} depend acutely on the estimate framework of flare energy, as a slight index change exponentially shifts the predicted frequencies of extreme events. Most notably, our energy–duration indices ($0.12\text{--}0.16$) challenge both magnetic reconnection theory predicting a energy-duration index near $\sim 1/3$ \citep{2015EP&S...67...59M, 2017ApJ...851...91N, 2024ApJ...975...69C} and self-organized criticality (SOC) theory with indices approaching $0.8$ \citep{2026ApJ...999L..28A}. This unresolved discrepancy highlights the complexity of the energy–duration relation and emphasizes the need for further investigation into the underlying physical mechanisms.

\section{Summary}\label{sect5}
Employing high-spatiotemporal-resolution SDO/HMI observations of 70 solar WLFs, we analyzed the true evolution of their source region’s radiating area and temperature. Based on this, we assess the physical plausibility of three previous simplified energy estimation methods of stellar flare and further propose a new energy estimation method that better matches the true evolution characteristics of solar WLFs. Finally, systematic comparison of energy estimates from these different methods was conducted. And the energy-related scaling laws of solar/stellar flares were also revisited. Our main results are listed as follows:

\begin{enumerate}
\item For most of the 70 solar WLFs, the area and temperature of their source regions undergo significant temporal evolution. The flare temperature evolves with the maximum amplitude of 28.2\%, much moderate compared to that of the flare area. These findings challenge the existing WLF energy estimation models assuming either a constant flare temperature or a fixed radiating area.

\item A new energy estimation method is proposed here that permits both flare area and temperature to evolve. Compared to the existing energy estimation models, our method calculates systematically smaller bolometric energy for the 70 solar WLFs. The median energy ratios of our method, \citet{2026ApJ...999L..18H}, and \citet{2024MNRAS.528.2562S} with respect to \citet{2013ApJS..209....5S} are 0.56, 1.15, and 1.50, respectively.

\item Revisiting scaling laws involved solar/stellar flare energy revealed that they could be systematically altered by the different energy estimation method choices. This finding highlights the need for great caution when comparing and interpreting energy-related statistical results derived under different energy estimation methods. And a re‑examination of these established statistical results and their targeted testing or revision are necessary in future work. 
\end{enumerate}

Finally, we would like to emphasize that for spatially-unresolved stellar flares, the existing energy estimation methods \citep{2013ApJS..209....5S, 2024MNRAS.528.2562S, 2026ApJ...999L..18H} have provided essential and remarkably effective approximations under observational limitations. To further reduce uncertainties in stellar flare energy estimates, future studies should explore the potential intrinsic or  empirical relationship between flare radiating area and temperature based on more high-resolution solar observations or simulations. This could achieve a parameter reduction that allowed us to solve for temperature and area simultaneously through the observed flare light curve. In addition, to resolve the remaining theoretical ambiguity between blackbody radiation and Hfb continuum emission mechanisms \citep{2013ApJS..209....5S, 2024MNRAS.528.2562S, 2024MNRAS.532L..56H, 2026ApJ...999L..18H}, future observational missions should prioritize high-sensitivity broadband spectrographs covering 3000--6000 {\AA} range to capture decisive spectral diagnostics such as the Balmer jump. In this context, the Solar UV--NIR Spectrometer (SUNS) developed by the Mackenzie Solar Observatory, now in the commissioning phase, will provide resolved solar spectra in the visible range, representing a step in this direction.

\section*{Acknowledgments}
The authors appreciate the anonymous referee for the constructive comments and valuable suggestions. The data used here are courtesy of the \emph{SDO} and \emph{GOES} science teams. Y.J.C. and Y.J.H. appreciate Prof. Heinzel, Prof. Sim{\~o}es, Dr. Bicz, Dr. Henggeng Han, Hengkai Ding, and Houle Huang for their insightful discussions and suggestions. The authors are supported by the Strategic Priority Research Program of CAS (XDB0560000), the National Key R\&D Program of China (2022YFF0503800), the National Natural Science Foundation of China (12273060, 12588202, and 12533010), the Youth Innovation Promotion Association CAS (2023063), China's Space Origins Exploration Program (GJ11020405), and the Specialized Research Fund for State Key Laboratory of Solar Activity and Space Weather. JFL also acknowledges support from the New Cornerstone Science Foundation through the New Cornerstone Investigator Program and the XPLORER PRIZE.

\bibliography{ref}{}
\bibliographystyle{aasjournal}
\end{document}